\documentclass[aps,prd,amssymb,cite,
amsfonts,epsf,preprintnumbers,nofootinbib,superscriptaddress]{revtex4}

\usepackage[dvips]{graphicx}
\usepackage{bm,latexsym,amsmath,amssymb,amsfonts}
\usepackage[usenames,dvipsnames]{color}
\usepackage[colorlinks=true,linkcolor=blue,citecolor=blue]{hyperref}
\usepackage{color}
\usepackage{ulem}
\usepackage{epsfig}
\usepackage{braket}
\usepackage[mathscr]{eucal}
\usepackage{cancel}
\usepackage{mathrsfs}
\usepackage{pgf, tikz}
\usepackage{slashed}
\usetikzlibrary{arrows,automata}

\definecolor{mypink1}{rgb}{0.858, 0.188, 0.478}
\definecolor{mypink2}{RGB}{219, 48, 122}
\definecolor{mypink3}{cmyk}{0, 0.7808, 0.4429, 0.1412}
\definecolor{mygray}{gray}{0.6}
\definecolor{pptbg}{rgb}{0.961,0.945,0.863}

\newcommand{\be}[1]{\begin{equation} \label{#1}}
\newcommand{\ee}{\end{equation}}
\newcommand{\bea}{\begin{eqnarray}}
\newcommand{\eea}{\end{eqnarray}}
\newcommand{\ba}{\begin{array}}
\newcommand{\ea}{\end{array}}
\newcommand{\nn}{\nonumber}

\newcommand{\bel}{\begin{align}}
\newcommand{\eel}{\end{align}}

\newcommand{\ve}[1]{\vec{\bm{#1}}}

\begin{document}
\title{Heat conduction in general relativity }

\author{Hyeong-Chan Kim and Youngone Lee}
\affiliation{School of Liberal Arts and Sciences, Korea National University of Transportation, Chungju 380-702, Korea}

\email{hckim@ut.ac.kr,  youngone@ut.ac.kr}


\begin{abstract}
We study the problem of heat conduction in general relativity by using Carter's variational formulation. 
We write the creation rates of the entropy and the particle as combinations of the vorticities of temperature and chemical potential.
We pay attention to the fact that there are two additional degrees of freedom in choosing the relativistic analog of Cattaneo equation for the parts binormal to the caloric and the number flows.
Including the contributions from the binormal parts, we 
find a {\it new} heat-flow equations
and discover their dynamical role in thermodynamic systems.
The benefit of introducing the binormal parts is that 
it allows room for a physical ansatz 
 for describing the whole evolution of the thermodynamic system.
Taking advantage of this platform, we propose a proper ansatz that deals with the binormal contributions starting from the physical properties of thermal equilibrium systems. 
We also consider the stability of a thermodynamic system in a flat background.
We find that {\it new} ``Klein" modes exist in addition to the known ones. 
We also find that the stability requirement is less stringent than those in the literature. 
\end{abstract}

\maketitle
 
\section{Introduction}

Heat conduction is crucial for us to study the thermal history of the universe and the quantitative modeling of collapsing objects into either a neutron star~\cite{Burrows1986,Riper1991,Aguilera2008,Cumming2017} or a black hole~\cite{Yuan2014,Shibata2005}.
We generally accept that an independent neutron superfluid permeates the inner crust of a neutron star.
The outer core contains superfluid neutrons, superconducting protons, and a highly degenerated gas of electrons. 
We still do not know how many independent fluids need to describe neutron stars having quark matter in the deep core~\cite{Alford:1999pb}.
The gravitational waves observed recent years~\cite{GW151226} come from two stars in a binary system. 
The heat conduction mechanism of dissipative fluids may account for its emission mechanism. 
These issues lead us to study the heat-conduction problem in general relativity. 

We adopt the variational approach of Taub~\cite{Taub54} and Carter~\cite{Carter72,Carter73,Carter89,Lopez2011,Andersson2007,Andersson:2013jga}.
In the approach, entropy is regarded as an independent field mathematically, like any other fluid elements.
The theory admits couplings between different fluid species in addition to self-interactions.
It makes the conjugate momentum not align with the velocity of a fluid, entrainment.
The existence of entrainment is crucial in a variety of physical situations~\cite{Andreev1975,Alpar1984,Andersson2010,Monsalvo2011,Andersson11}. 
Entrainment between the caloric and the number flows in a system is the essential ingredient that allows one to derive a Cattaneo-type form of Fourier's law and thus restore causality to heat conduction within the variational approach~\cite{Andersson2011}. 
It is also crucial in curing ~\cite{Priou1991} the instability~\cite{Olson1990} in the original theory of Carter.
The model successfully resolves troublesome issues associated with causality and leads to the emergence of the expected second sound~\cite{Priou1991,Carter1994}. 
The resulting model is equivalent to the Israel-Stewart theory~\cite{Andersson2011}.
For a comprehensive review on variational thermodynamics, see Ref.~\cite{AnderssonNew}. 

Let us address the causality problem of heat conduction in Newtonian physics and general relativity.
In Newtonian physics, the heat equation states that a temperature distribution $\Theta$ evolves as 
\be{T eq}
\frac{\partial \Theta}{\partial t} = - \frac{1}{c_V} \nabla\cdot \ve{q},
\ee
where $c_V$ is the volumetric heat capacity. 
The heat flux $\ve{q}$ is, in turn, given by Fourier's law:
\be{Q eq}
\ve{q} = - \kappa \nabla \Theta,
\ee
where $\kappa\geq 0$ is the thermal conductivity. 
Putting Eq.~\eqref{Q eq} into Eq.~\eqref{T eq}, one may notice that the resulting heat equation is parabolic.
This fact implies an undesirable property: information propagates instantaneously.
The efforts to rectify this deficiency had led to Cattaneo equation, 
\be{C eq}
\ve{q}+\tau_R \frac{\partial \ve{q}}{\partial t}  = - \kappa \nabla \Theta,
\ee
where $\tau_R$ is some small positive number, 
that is a relaxation time-scale for the medium~\cite{Herrera1997}.
In this equation, one has restored the causality through a time-dependent term at the cost of introducing a term that does not come from underlying microphysics.

If one applies the heat equation to general relativity, the causality problem becomes more severe because no information may travel faster than the speed of light.
Israel \& Stewart~\cite{IS1,IS2,IS3} resolved the problem of heat propagation for the first time.
They postulated the entropy flux as a series of terms that encode deviations from thermal equilibrium. 
The theory is thus rather complex because the coefficients of this expansion need to be determined phenomenologically. 
However, it allows one to recover a causal heat equation, a relativistic analog of Cattaneo equation.
 
One usually obtain Cattaneo-like equation in the following way.
The time part of the energy-momentum conservation equation, $\nabla_aT^{ab}=0$, can be written as a formula for the entropy creation rate,
\be{Gs H}
\Gamma_s \equiv \nabla_a s^a = \frac{q^a H_a}{\Theta^2} ,
\ee
where  $H_a$ is a spatial vector orthogonal to the particle trajectory and consists of thermodynamic quantities such as the temperature gradient, $\nabla_a \Theta$.
Here, the conservation of the particle number is assumed implicitly.  
To inscribe the second law of thermodynamics, one usually use the easiest choice to make the creation rate have a quadratic form for the heat, $\Gamma_s \propto q^2$, so that the creation rate is non-negative.
Therefore, the relativistic analog of the Cattaneo equation takes the form:
$$
 q^a = \kappa \gamma^{ab}H_b,
$$ 
where $\gamma^{ab}$ is the projection operator normal to the particle trajectory.
However, as briefly mentioned in Ref.~\cite{Andersson2011}\footnote{The authors mentioned this fact in terms of the matter and the caloric forces rather than the $Q_a^\perp$ vectors.}, this choice of $q^a$ is not a general solution to the problem.
One may notice that any addition of $Q_a^\perp$, normal to both the particle trajectory and $q^a$, to the right-hand side of this equation does not hurt the argument. 
Therefore, there are two additional degrees of freedom in the choice of $q^a$.

In this work, we pursue this possibility thoroughly.
During the course, we show that the ansatzes in the literature correspond to different choices of the binormal field $Q^\perp_a$.
Remember that `regular' ansatz of Carter~\cite{Carter89} leads to instability of the system.
On the other hand, Lopez-Monsalvo \& Andersson's choice in Ref.~\cite{Andersson2011} leaves no instability.
This fact tells that the $Q^\perp_a$ field is not a kind of gauge choice but significantly affects the physical properties of a thermal system.
Therefore, an appropriate choice of them is crucial in understanding the physical behaviors of a thermodynamical system.
We suggest a natural choice based on the physics in thermal equilibrium configuration.

\vspace{.2cm}
We begin our analysis by summarizing the original variational approach for two fluids in Sec.~\ref{sec:2}. 
Then, we reformulate the relativistic heat conduction equations in Sec.~\ref{sec:3}.
We introduce binormal vectors also.  
In Sec.~\ref{sec:4}, we analyze the thermal equilibrium condition and write Cattaneo equation in a general form.
Then, we propose a conjecture on how to determine the incomplete parts of the equation of motion by using the information on thermal equilibrium. 
In Sec.~\ref{sec:5}, using the choice for the binormal fields $Q^\perp_a$ and $\tilde Q^\perp_a$, we consider linear perturbations around a thermal equilibrium configuration in a flat Minkowski background.
In Sec.~\ref{sec:6}, we summarize the results.
Three appendices support the calculations in this work. 

\section{The two-constituent model for relativistic thermodynamics} \label{sec:2}
In this section, we survey the two-constituent model of Carter~\cite{Carter89} for thermodynamics. 
The particle number in the system is assumed to be large enough so that the fluid approximation applies and there is a well-defined matter current $n^a$.  
As discussed in~\cite{Haskell2012}, this is the same as assuming that each constituent has a short enough internal length scale to perform averaging, while any mechanism that couples the flows acts on a larger length scale or a longer time scale. 
A typical system of this kind is laboratory superfluids~\cite{Carter94,Andersson11}.
In this model, one treats entropy as a fluid with a flux $s^a$.
This current is, in general, not aligned with the particle flux $n^a$. 
The misalignment is associated with the heat flux, $q^a$, and leads to entropy creation. 

\vspace{.1cm}
The intrinsic elegance of an action principle is that once an ``equation of state'' for matter, here the master function $\Lambda$, is given, the theory provides the relation between the various currents and their conjugates.
The scalar $\Lambda$ is a function of different scalars that the two fluxes $s^a$ and $n^a$ can form: 
\be{sxn}
|s| = (-s_a s^a)^{1/2}, \qquad |x| = (- s_a n^a)^{1/2}, \qquad |n| = (-n_a n^a)^{1/2},
\ee
where the inequality,
\be{inequality}
|x|^2-|s|\, |n| \geq 0,
\ee 
is saturated when the fluid is in thermal equilibrium.
Later we write $n\equiv|n|$ and $x\equiv |x|$. 
However, $s \neq |s|$.  
The Eulerian variation of $\Lambda$ for the three scalars is
\be{dl Lambda}
\delta \Lambda = \Theta_a \delta s^a + \chi_a \delta n^a + \frac12 \left( \Theta^a s^b+ \chi^a n^b \right) \delta g_{ab} ,
\ee
where $\Theta_a$ and $\chi_a$ are the conjugates to $s^a$ and $n^a$, respectively.
Because the partial differentiations commute with each other, 
there exists a symmetry written as
\be{sym}
\Theta_{[a} s_{b]} + \chi_{[a} n_{b]} = 0.
\ee
For a single matter fluid, one usually imposes the conservation of the particle number,
\be{nabla n}
\Gamma_n \equiv \nabla_a n^a =0.
\ee
However, we keep $\Gamma_n$ alive formally during the formulation procedure and impose the constraint in the final construction as done in Ref.~\cite{Carter89}.

\vspace{.2cm}
In this work, we adopt the matter (Eckart) frame, in which a particle follows an integral curve generated by a vector field $u^a$ so that
\be{u:def}
u^a = \frac{n^a}{n}.
\ee
With this choice, we have $n = |n|$ and $x = |x|$ naturally. 
On the other hand, a scalar entropy density flowing with the matter is 
$
s \equiv -u_a s^a\neq |s|
$, 
where $s/|s|$ is the redshift associated with the relative motion of matter and entropy frames.
To proceed, we introduce an additional vector field $q^a$, which denotes the deviation of the caloric flow from that of the matter flow. 
By definition, the vector is orthogonal to $u^a$ and satisfies
\be{s:thermal1} 
s^a =  s u^a + \frac{q^a}{\Theta},   \qquad q^a u_a =0,\qquad 
 s =\frac{x^2}n,
\ee
where $\Theta \equiv -u^a\Theta_a$ and $ q/\Theta = \sqrt{(x^2/n)^2-|s|^2}$.
The inequality~\eqref{inequality} constrains that $q^a$ is spacelike or null.
We now have $8$-independent unknown parameters $n$, $s$, $u^a$, and $q^a$.

Let us denote the conjugate covectors in this frame with the form:
\be{bar chi}
\Theta_a  
=  \Theta u_a + \vartheta_a , 
  \qquad \vartheta_a \equiv  \beta q_a,  \qquad 
\chi_a = \mu \Theta_a + \nu q_a = \mu \Theta u_a 
	+\alpha q_a; \qquad \alpha \equiv  \mu\beta +\nu,
\ee
where $\Theta$ and $\chi$ are the temperature and the chemical potential measured by a comoving observer with the matter, respectively. 
The symmetry~\eqref{sym} presents a relation between the parameters,
\be{ABC}
n \alpha + s\beta = 1 
\quad \rightarrow \quad \alpha = \frac{1- s\beta}n .
\ee
From the reality of the size of the number flow, $|\bm{\chi}|= \sqrt{-\chi_a \chi^a} $, one gets an upper bound of the heat:
$$
q^2 \leq \frac{\mu \Theta^2}{\alpha}.
$$
Note that all the momenta lie in a plane made of $u^a$ and $q^a$.

\vspace{.2cm}

The integral curves of $n^a$ and $s^a$ compose two congruences of worldlines. 
Let us consider variations of the worldlines caused by the displacements for $s^a$ and $n^a$ determined by the vector fields $\xi_0^a$ and $\xi_1^a$, respectively.
We also consider the Eulerian variation of the metric $\delta g_{ab} $.
The variations of $n^a$ and $s^a$ become~\cite{Carter89},
\be{delta s n}
\delta s^a =  - (\nabla \cdot \xi_0) s^a + [s, \xi_0]^a-\frac12(g^{cd} \delta g_{cd}) s^a  , \quad
\delta n^a =  - (\nabla \cdot \xi_1) n^a + [n, \xi_1]^a-\frac12(g^{cd} \delta g_{cd}) n^a  .
\ee
The corresponding action is
\be{S}
I = \int_{\mathcal{M}} d\mathcal{M} \Lambda(|s|,|x|,|n|) ,
\ee
where $d\mathcal{M}$ represents the volume measure over the 4-dimensional manifold. 
After varying Eq.~\eqref{S}, one finds~\cite{Carter89}
\be{d I}
\delta I = \int_{\mathcal{M}} \bm{d} \mathcal{M}: (-f^0_a \xi_0^a 
	-f^1_a \xi^a_1 
		+\frac12 T^{ab} \delta g_{ab}) 
		+ 2 \oint_{\partial \mathcal{M}} \,\bm{d}\Sigma_a
	(s^{[a} \xi_0^{b]} \Theta_b + n^{[a} \xi_1^{b]} \chi_b ) 
	.
\ee
Here, the stress tensor is, explicitly,   
\be{T:2C}
T_{ab} = \Theta_a s_b + \chi_a n_b + \Psi g_{ab} 
=  \left( \frac{\beta}{\Theta} q^2 
	- \Lambda \right) u_a u_b 
	+ 2 u_{(a} q_{b)} 
	+\Psi \gamma_{ab} 
	+\frac{\beta}{\Theta}q_{(a} q_{b)},
\ee
where we use the symmetry relation~\eqref{sym}, and $\Psi$ denotes the pressure
\be{bar Psi}
\Psi = \Lambda - \Theta_a s^a - \chi_a n^a .
\ee
From this stress tensor, one finds the heat flow covector has its desired form:
$q_a= -\gamma_{ab} T^{bc} u_c $, where 
\be{gamma}
\gamma_{ab} = g_{ab} +u_a u_b
\ee
denotes the projection operator normal to $u^a$.
The forces are, by using the differential form notation,  
\be{forces}
\bm{f}^0 = \bm{\Theta} (\nabla \cdot \ve{s} ) 
	+ \ve{s} \cdot (\bm{d \Theta}), \qquad
\bm{f}^1 = \bm{\chi} (\nabla \cdot \ve{n} ) + \ve{n} \cdot (\bm{d \chi}).
\ee
Here $[\bm{d \Theta}]_{ab} = 2\nabla_{[a} \Theta_{b]}$ and 
	$[\bm{d\chi}]_{ab} = 2\nabla_{[a} \chi_{b]}$.
The energy-momentum conservation relation, $\nabla_a T^{ab}=0$, for an isolated fluid becomes
\be{bianchi}
\nabla_a T_b^{a} = f^0_b + f^1_b =0. 
\ee

Interpreting the master function, $\Lambda$, as a function of $s$, $x$, and $n$ and using Eq.~\eqref{s:thermal1}, we have the variational relation for $\Lambda$ to be $\delta \Lambda = - \chi \delta n - \Theta \delta  s + \vartheta d(q/\Theta)$.
Now, we write the energy density and the pressure from the master function,
\bea \label{rho:2C}
\rho &=& u^au^bT_{ab} =\frac{\vartheta q}{\Theta}  - \Lambda , \\
\Psi &=& \Lambda-\Theta_a s^a -\chi_a n^a 
= \Lambda+ \Theta s + n \chi - \vartheta\frac{q}{\Theta} = \Theta s + n \chi - \rho . \label{Psi}
\eea
We can use Eq.~\eqref{rho:2C} as a Legendre transform from $\Lambda (n,s,q/\Theta)$ to $\rho(n,s,\vartheta)$ using $\vartheta$ as a parameter rather than $q/\Theta$,
\be{d rho}
\rho(n,s,\vartheta) = \frac{q}{\Theta} \vartheta -\Lambda \quad \Rightarrow \quad
	d\rho = \Theta ds+\chi dn + \frac{q}{\Theta} d\vartheta .
\ee
Now, interpreting $\rho$ as a function of $n$, $s$, and $p$, the conjugates are
\be{conj}
\Theta(n,s,\vartheta) = \left(\frac{\partial \rho}{\partial s}\right)_{n,\vartheta}, \qquad
\chi(n,s,\vartheta) = \left(\frac{\partial \rho}{\partial n}\right)_{s,\vartheta}, \qquad 
q(n,s,\vartheta)= \Theta \left(\frac{\partial \rho}{\partial \vartheta}\right)_{n,s}.
\ee
These equations allow us to interpret $\Theta$ and $\chi$ as the temperature and  the chemical potential, respectively.

\section{The heat-conduction equations and the binormal quantities} \label{sec:3}

In this section, we reformulate the relativistic analog of Cattaneo equation, which was studied originally by Carter~\cite{Carter89}
 and improved later by Priou~\cite{Priou1991} and Lopez-Monsalvo \& Andersson~\cite{Andersson11}.
As the authors already mentioned in their works, they had chosen the simplest way of their own  to achieve the entropy creation to be quadratic in the source or to get other benefit.
Their choices had resulted in separate equations for the same system, one has instability and the other does not.
We show that both are parts of a general equation for specific choices of the binormal vector $Q_a^\perp$. 

\subsection{Creation rates}

To formulate an improved heat conduction equation, we use i) the formal symmetry of the formulation between $n^a$ and $s^a$. 
 ii) The physics must be independent of the choice of observers who follow $n^a$ and $s^a$.

\vspace{.2cm} 
The energy conservation law, $\nabla_a T^{a}_b =0$, with relations~\eqref{forces} and \eqref{bianchi} gives the creation rate of the entropy density in two ways:
\bea
\label{udT}
\ve{u} \cdot( \bm{f}^0+ \bm{f}^1)=0 \quad &\rightarrow& \quad
\Gamma_s +\left(\frac{\chi}{\Theta}\right) \Gamma_n 
	=  -\frac{q^a ( \ve{u} \cdot \bm{d \Theta})_a}{ \Theta^2 }, \\
\label{sdT}
\ve{s} \cdot( \bm{f}^0+ \bm{f}^1)=0 \quad &\rightarrow& \quad
\Gamma_s +\left(\frac{\bm \chi\cdot \bf s}{\bf \Theta\cdot s}\right) \Gamma_n
	=  \frac{q^a ( \ve{u} \cdot \bm{d \chi})_a}{\sigma_\Theta \Theta^2 },
	~~~~~~\sigma_\Theta \equiv \frac{\ve{s}\cdot\bm{\Theta}}{\ve{n}\cdot\bm{\Theta}},
\eea
where $\sigma_\Theta$
is the specific entropy measured by an observer which follows the temperature covector.
In the absence of  heat, the temperature covector is parallel to $n_a$, making $\sigma_\Theta  \to  \sigma=s/n$.
\\
We interpret these two equations as  temperature, 
chemical potential, and heat create the entropy and the particles.
When the heat does not flow, $\ve{s}\parallel\ve{u}$, these two take the same form.
Else the two equations give the two creation rates as functions of $\Theta^a,~\chi^a$ and $q_a$:
\bea
\label{entropy gen}
~~~~~~~~~~~~~~~~~~\Gamma_s &=&\frac{1-\gamma}{ \gamma} \frac{q^a}{\Theta^2}
                         \left(	\ve{u} \cdot \bm{d \Theta} 	+\frac{\ve{u}\cdot \bm{d \chi}}{\sigma_\chi}\right)_a ,
	~~~~~~~~~~\sigma_\chi \equiv \frac{\ve{s}\cdot\bm{\chi} }{\ve{n}\cdot \bm{\chi}},
       ~~~\gamma \equiv  1- \frac{\sigma_\chi}{\sigma_\Theta},
\eea
where $\sigma_\chi$ is the specific entropy similar to $\sigma_\Theta$ for the chemical potential covector, and
\bea
\label{number gen}
\Gamma_n &=& -\frac{q^a }{\gamma\chi\Theta} 
    \left( \ve{u} \cdot \bm{d \Theta}  +\frac{  \ve{u} \cdot \bm{d \chi} }{\sigma_\Theta}\right)_a .
       ~~~~~~~~~~~~~~~~~~~~~~~~~~
\eea
Note that both the caloric and the number flows contribute to the entropy and the particle creations. 
Contrary to the previous literature in which $\Gamma_n=0$ from the beginning, we keep  $\Gamma_n$ alive. 
Our approach will be appropriate even when $\Gamma_n\neq 0$.
We will show the consistency for the limit $\Gamma_n\to 0$ later.

Since $1-\gamma=\sigma_\chi/\sigma_\Theta$,
we get $\gamma<1$ because the term $\sigma_\chi/\sigma_\Theta$ is positive definite when all the vectors are future directed timelike vectors.
Note that $\gamma$ vanishes when the specific entropy measured by a comoving observer with $\Theta_a$ is identical to that with $\chi_a$, i.e., $\sigma_\Theta=\sigma_\chi$:
\be{LL}
\frac{\bm s\cdot\bm \Theta}{\bm n\cdot\bm\Theta} =\frac{\bm s\cdot\bm \chi}{\bm n\cdot\bm \chi}  \quad \rightarrow \quad \Theta_c \parallel \chi_c
~~\mbox{or} ~~{\bm\chi}=\mu{\bm\Theta}.
\ee 
Explicitly, 
$$
\gamma = \frac{\nu}{s} \frac{q^2}{\Theta \chi} 
	\left[1- \frac{\beta q^2}{s \Theta^2}\right]^{-1}
$$
vanishes in the limit $q\to 0$ in addition to the $\nu \to 0$ limit keeping $q$ finite, where this second limit is the Landau-Lifshitz model~\cite{LL}.  
Therefore, one should be careful in taking the $\gamma \to 0$ limit. 
From now on, we adopt the notations $\bar\sigma\equiv\sigma_\Theta$ and $\sigma_\chi=(1-\gamma)\bar\sigma$.

\subsection{New heat-flow equation}
The second law of thermodynamics dictates that the creation rate of the entropy should be non-negative for an isolated system~\cite{Chamel, Hiscock}.
To impose the second law of thermodynamics, one needs to make the creation rate $\Gamma_s$ in Eq.~\eqref{entropy gen} 
a quadratic function of the heat flow $q^a$, $\Gamma_s \propto q^2$.
For this purpose, we write the heat-flow vector to the form:
\be{newheat}
q_a \equiv \kappa\frac{1-\gamma}{\gamma}  \left(
	\ve{u} \cdot \bm{d \Theta} 
  +\frac{\ve{u}\cdot \bm{d \chi}}{\bar{\sigma}(1-\gamma)}
	 \right)_a
	    + Q^\perp_a ,
\qquad Q^\perp_a = \perp_a^b Q^\perp_b,
\ee  
where $\kappa$ denotes the heat conductivity, and
\be{perp}
\perp_a^b \equiv \delta_a^{b}+u_au^b - \frac{q_aq^b}{q^2} 
\ee
represents the projection operator to a 2-dimensional space $\Sigma_\perp$ binormal to $u^a$ and $q^a$ both.
The importance of this plane $\Sigma_\perp$ lies in the fact that no heat flows along with it, which reminds us of the thermal equilibrium state.
We will call this equation {\it the relativistic Cattaneo equation}.

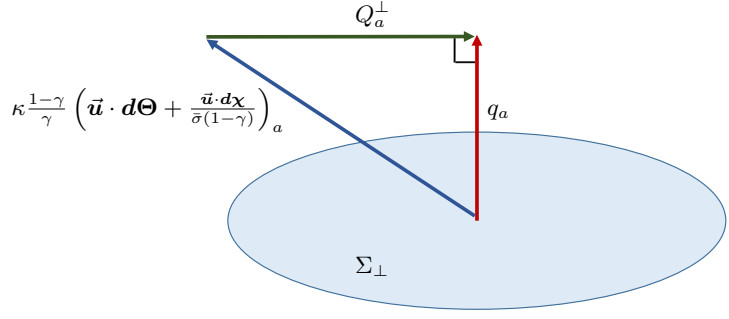
\begin{figure}[thb]
\begin{center}
\includegraphics[width=0.42\linewidth,origin=tl]{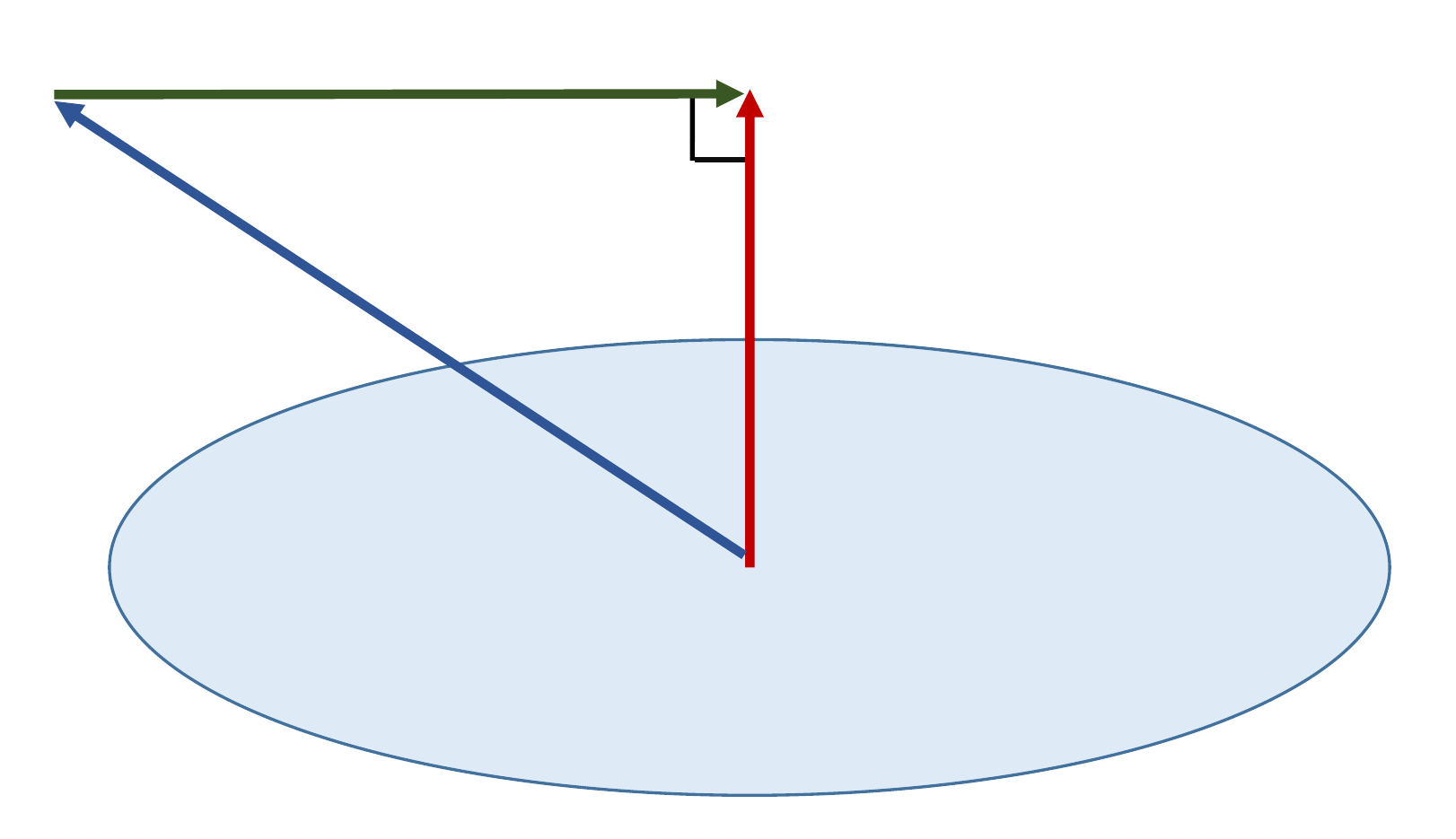} 
\put(-280,80){$ {\kappa} \frac{1-\gamma}{\gamma} \left(
	\ve{u} \cdot \bm{d \Theta}
	+\frac{\ve{u}\cdot \bm{d \chi}}{\bar{\sigma}(1-\gamma)}
  \right)_a$}
  \put(-150,115){$Q_a^\perp$}
  \put(-100,80){$q_a$}
  \put(-150,20){$\Sigma_\perp$}
\end{center}

\caption{Graphical definitions of $Q_a^\perp$ and $\kappa$.
$\Sigma_\perp$ is the 2-dimensional space perpendicular to both $u^a$ and $q^a$.
}
\label{Qq}
\end{figure}

Note that this definition of $Q_a^\perp$ also defines $\kappa$ so that $Q_a^\perp$ is binormal (see Fig.~\ref{Qq}).
Given $\ve{u}$ and $\bm{\Theta}$, one can obtain $Q_a^\perp$ to be
\be{Q: Theta}
Q_a^\perp ~= -{\kappa} \frac{1-\gamma}{\gamma} \perp_a^c\left(
	\ve{u} \cdot \bm{d \Theta}
	+\frac{\ve{u}\cdot \bm{d \chi}}{\bar{\sigma}(1-\gamma)}
  \right)_c,
\ee
by applying $\perp_a^b$ to Eq.~\eqref{newheat}.

The equation~\eqref{entropy gen} and the relativistic Cattaneo equation~\eqref{newheat} give a non-negative entropy creation rate,
\be{2nd law}
\Gamma_s = \frac{q^2}{\kappa \Theta^2} \geq 0.
\ee
Note that the first term of the relativistic Cattaneo equation in Eq.~\eqref{newheat} guarantees the non-negativeness.
The additional  $Q^\perp_a$  term plays no role here.

Considering the particle creation rate~\eqref{number gen}, 
one obtains the following expression by taking into account another binormal term $\tilde Q_a^\perp$,
\be{vorti:Gamma n}
 ( \ve{u} \cdot \bm{d \Theta})_a
 	+ \frac{1}{\bar\sigma}    ( \ve{u} \cdot \bm{d \chi})_a 
	= \gamma \left( \tilde{Q}^\perp_a  -\gamma_n\frac{q_a}{\kappa} \right),
\ee
where $\tilde Q^\perp_a = \perp_a^b \tilde Q^\perp_b$ is also a binormal vector, and we have introduced a dimensionless quantity for the particle creation rate, 
\be{gamma n}
\gamma_n \equiv  \frac{\kappa \Theta \chi \Gamma_n}{q^2} .
\ee
This equation reproduces Eq.~\eqref{number gen} because $q^a\tilde Q_a^\perp=0$.
Therefore, it is equivalent to Eq.~\eqref{number gen} except for the two additional equations, which compensate for the addition of the auxiliary  vector $\tilde Q^\perp_a$.
The creation rate may vanish even in the presence of  heat.
In this case, $\Gamma_n =0$, we have 
\be{const:Q'}
 ( \ve{u} \cdot \bm{d \Theta})_a
 	+\frac{1}{\bar{\sigma}}  ( \ve{u} \cdot \bm{d \chi})_a 
	= \gamma \tilde{Q}^\perp_a .
\ee   
One may notice that the $\gamma \to 0 $ limit of $\Gamma_s$ in Eq.~\eqref{entropy gen}  behaves well with the support of this equation.

Now putting Eq.~\eqref{vorti:Gamma n} to Eq.~\eqref{newheat}, we write the heat $q_a$ in a more familiar form.
We get the relativistic analogue of Cattaneo equation in a similar form in the literature up to the binormal fields,
\be{eq1}
\left(1 +\gamma_n\right) q_a = - \kappa ( \ve{u}\cdot \bm{d\Theta})_a  + Q^\perp_a +\kappa \tilde{Q}^\perp_a  
 .
\ee
In terms of the matter flow, the above equation is expressed as, by using Eq.~\eqref{vorti:Gamma n}, 
\be{eq2}
\left[1 +  (1-\gamma)\gamma_n \right]q_a =  \frac{\kappa}{ \bar \sigma} (\ve{ u} \cdot \bm{d \chi})_a+ Q^\perp_a + \kappa (1-\gamma) \tilde{Q}^\perp_a .
\ee
The two equations~\eqref{eq1} and \eqref{eq2} are equivalent to the two equations~\eqref{newheat} and \eqref{vorti:Gamma n}.
Therefore, we may use these two equations instead of the previous two as independent relativistic heat-flow equations.
To distinguish the two equations, we call the first~\eqref{eq1} the relativistic analog of Cattaneo equation and the second~\eqref{eq2} the matter-flow equation. 
When we call both, we use the terminology ``heat-flow equations".

Finally,  the binormal directions of the equation $\bm{f}^0+\bm{f}^1=0$ present
\be{perp}
 (\sigma - \bar{\sigma}) \frac{Q^\perp_a}{\kappa} 
+ \left[\sigma - (1-\gamma)\bar\sigma\right] \tilde{Q}^\perp_a 
=-\frac{1}{n\Theta}\perp_a^c\left[ \ve{q} \cdot \bm{d\Theta} \right]_c .
\ee  
Here, $\sigma - \bar\sigma = \frac{q^c\Theta_c}{n \Theta^2}=\frac{\beta q^2}{n \Theta^2} $.

Even though the presence of this $Q_a^\perp$ term has been known implicitly in previous literature,
 its dynamical role has not been addressed explicitly.
This is because no physical principle was known to determine them. 
In general, one may introduce higher-order terms of $q_a$ to Eq.~\eqref{newheat}, which possibility we does not pursue in this work.

\section{Local thermal equilibrium} \label{sec:4}

Let us count the degrees of freedom. 
We have twelve-independent parameters to be determined, the eight-dynamical variables $n$, $s$, $u^a$, $q^a$, and the four auxiliary variables, $Q^\perp_a$, and $\tilde{Q}^\perp_a$. 
On the other hand, we have ten equations of motion,
one from the particle creation rates Eq.~\eqref{nabla n} or something similar, one from the entropy creation rate~\eqref{2nd law}, three from the relativistic analog of Cattaneo equation~\eqref{eq1}, 
three from the matter-flow equation~\eqref{eq2}, and lastly, two from the binormal part equation of the energy-momentum conservation equation~\eqref{perp}.
Therefore, we need two additional equations to determine the whole evolution of a thermodynamic system.

Here, we examine which part of the equations of motion are lost and how to compensate for those missing based on a physically sound conjecture.
We make the conjecture based on the thermodynamic equilibrium configuration.

\subsection{Thermodynamic equilibrium configurations}
First, we consider a thermal equilibrium state with equilibrium values $\rho$, $n$, $u^a$, $\mu$, $\Theta$, and $g_{ab}$.
In the equilibrium, there is no heat $q=0$.
Now, we display the equations satisfied by those equilibrium parameters one by one.
The analysis follows the recent work~\cite{Kim:2021kou}.
\begin{enumerate}
\item In thermal equilibrium, the particle number must be conserved:
\be{dn}
\Gamma_n= \dot n+ n \nabla_a u^a = 0.
\ee
\item The second law of thermodynamics~\eqref{2nd law}, with $q=0$, gives 
\be{ds}
\Gamma_s 
	= \dot s + s \nabla_a u^a = s\frac{d}{d\tau}\log \sigma=0,
\ee
where we use Eq.~\eqref{dn} in the second equality. 
Note that the entropy conservation constrains the specific entropy $\sigma=s/n$ to be constant in time for an equilibrium state.

\item Finally, the relativistic Cattaneo equation in the absence of heat gives
\be{Tolman Klein}
 \gamma^{ab} \mathcal{T}_a =0 ,
 \ee
where, for later convenience, we define a parameter 
\be{T}
\mathcal{T}_a \equiv \Theta \dot u_{a} + \nabla_a \Theta, \qquad \dot u_a \equiv u^c \nabla_c u_a.
\ee
For an appropriate choice of the geometry, the equation gives Tolman relation for the temperature in thermal equilibrium~\cite{Tolman,Santiago:2018lcy,Kim:2021kou}.

\item 
The matter flow equation in combination with the relativistic analog of Cattaneo equation gives Klein's relation
~\cite{Klein49} for a single number flow in thermal equilibrium, 
\be{Klein}
\gamma_b^c \nabla_c \mu =0.
\ee
\end{enumerate}
As shown in Ref.~\cite{Lima:2019brf}, the Tolman's and the Klein's relations, combined with the definition of  pressure~\eqref{Psi}, reproduce the energy-momentum conservation equation, $\nabla_a T^{ab} =0$. 
Therefore, we do not need to examine the energy-momentum conservation law.
In the analysis, we assume $Q^\perp_a = 0 = \tilde{Q}^\perp_a $ in the equilibrium state. 
This is natural because one cannot find the binormal directions in the absence of a heat.

\subsection{The heat-flow equations and local equilibrium}
To find a consistent conjecture which compensates for the lost equations, we first derive a relation satisfied by the additionally introduced binormal fields, $Q^\perp_a$, $\tilde{Q}^\perp_a$, and the sources for the heat.

In appendix~\ref{app:constrain}, we combine the two heat-flow equations~\eqref{eq1} and \eqref{eq2} by using the relation~\eqref{bar chi}.
We then project the equation to the binormal direction by multiplying $\perp_a^c$ to get  
\be{d mu eq}
 \alpha \left( \frac{Q^\perp_a}{\kappa}   
 	+\tilde{Q}^\perp_a\right) 
 + \beta \bar\sigma \left(\frac{Q^\perp_a}{\kappa} 
 	+(1-\gamma) \tilde{Q}^\perp_a\right) 
= -\perp_a^c\left[\beta \Theta  (\nabla_c \mu)  
	-\nu \mathcal{T}_c 
 \right]
.
\ee
Note Eq.~\eqref{perp} to see another combination of the binormal vectors,
$\sigma( Q^\perp_c/\kappa   +\tilde{Q}^\perp_c) 
	- \bar\sigma [Q^\perp_c/\kappa +(1-\gamma) \tilde{Q}^\perp_c].
$

Combining the binormal direction components equation~\eqref{perp} with Eq.~\eqref{d mu eq}, one gets the binormal vectors.
Before going further, we first write down the general form for the binormal vectors.
Rather than writing down the individual components for the binormal vectors, it is better to write down the following combinations: 
\bea 
\frac{Q^\perp_a}{\kappa} + \tilde{Q}^\perp_a 
&=& -\perp_a^b \left( 
     \frac{\beta}{\Theta} [\ve{q}\cdot \bm{d\Theta}]_b 
 	+ n\beta \Theta  \nabla_b \mu 
	-n\nu \mathcal{T}_b\right) , \nn \\
\frac{Q^\perp_a}{\kappa} + (1-\gamma) \tilde{Q}^\perp_a  &=&  
 \frac{1} { \bar\sigma} \perp_a^b
 \left(\frac{\alpha}{\Theta} [\ve{q}\cdot \bm{d\Theta}]_b 
		-\sigma n \beta \Theta \nabla_b \mu 
		+ \sigma n\nu \mathcal{T}_b  
\right), \label {Qs}
\eea
because they take simpler forms and take part in the equations~\eqref{eq1} and \eqref{eq2}.
Here, we use $\alpha +\beta \sigma = 1/n$ in Eq.~\eqref{ABC}.
We now know the binormal vectors in terms of the binormal parts of $[\ve{q}\cdot \bm{d\Theta}]_a$, $\nabla_b \mu$, and $\mathcal{T}_a$.

Putting these binormal vectors, the relativistic analog of Cattaneo equation~\eqref{eq1} becomes
\be{eq1-2}
\frac{q_a}{\tilde\kappa} 
    + \beta \left[ \dot q_a + (\nabla_a u_b) q^b +\frac{ 
      \perp_a^c \left[ \ve{q}\cdot \bm{d\Theta}\right]_c 
}{ \Theta} \right] 
= -\frac{q_aq^b}{q^2}\mathcal{T}_b
-  \perp_a^bn\beta\left[ (\mu+\sigma) \mathcal{T}_b +  \Theta \nabla_b \mu\right]  ,
\ee
where, the $ [\ve{q}\cdot \bm{d\Theta}]_b $ term can be simplified by using Eq.~\eqref{q d Theta} and 
\be{tildes}
\tilde \kappa \equiv \frac{\kappa}{ 1+\gamma_n +\kappa \dot \beta} .
\ee
The equation shows a few aspects compared to the corresponding equation (15.51) in Ref.~\cite{AnderssonNew}:
\begin{enumerate}
\item They differ only by the terms including $\perp_a^c$ on both sides.
These terms represent the binormal dependence of the temperature gradient and acceleration.
For a thermal system, when the right-hand side of the first equation of Eq.~\eqref{Qs} vanishes, $\perp_a^b \left( 
     \frac{\beta}{\Theta} [\ve{q}\cdot \bm{d\Theta}]_b 
 	+ n\beta \Theta  \nabla_b \mu 
	-n\nu \mathcal{T}_b\right)=0 $, the equation exactly reproduces the one in Ref.~\cite{AnderssonNew}.
Because this choice is optional, the present formulation for Cattaneo equation is more general.

\item First-order deviations from equilibrium with the same conditions $\nabla_au_b=0$ and $\gamma_n=0$ with the reference~\cite{Olson1990} give the same relaxation time scale, $\tau_R=\kappa\beta/(1+\kappa\dot\beta)$.
In the presence of  particle creation, the time scale changes to $\tau_R = \kappa \beta/(1+ \gamma_n+\kappa \dot \beta)$.

\item There are nonlinear corrections in the $[\ve{q}\cdot \bm{d\Theta}]_c$ terms. 
The nonlinear term affects only the binormal directions.

\item The contribution of the source term on the right-hand side consists of two parts.
One is the $q^a$ direction because of the Tolman temperature gradient $\mathcal{T}_b$, and the other is the binormal contributions.

\end{enumerate}

Finally, the remaining matter-flow equation~\eqref{eq2} becomes
\bea	\label{matter-flow}
&&\frac{q_a}{\kappa'} - \frac{\alpha}{\bar\sigma}
	\left[ \dot q_a + (\nabla_a u_b) q^b
		+ \perp_a^b\frac{ 
		  [\ve{q}\cdot \bm{d\Theta}]_b } 	
 		{ \Theta}\right]  \nn\\
&& \qquad
= \frac{q_aq^b}{q^2} \frac{1}{\bar\sigma}\left[\mu\mathcal{T}_b
	 +  \Theta\nabla_b\mu \right]    
	+ \perp_a^b\frac{ n\alpha}{\bar\sigma }
	\left[(\mu+\sigma) \mathcal{T}_b+ \Theta \nabla_b \mu  \right] ,
\eea
where 
\be{kappa'}
\kappa' \equiv \kappa \left[ (1+(1-\gamma)\gamma_n )
	-\frac{\kappa  \dot\alpha}{\bar\sigma}
    \right]^{-1}. 
\ee
The $q^a$ direction part of this equation is identical to the $\bm{f}^1$ equation taking its right-hand side to be Eq.~\eqref{f1 comp}.
Because now all of the binormal vector fields are replaced by the $[\ve{q}\cdot \bm{d \Theta}]_a, \nabla_a \mu$ and $\mathcal{T}_a$, we have 8-independent parameters. 
We also have 8-differential equations to determine them, \eqref{eq1-2}(3), \eqref{matter-flow}(3), the particle number conservation (1), and the entropy creation equation~\eqref{2nd law} (1). 
Since the binormal parts of equations~\eqref{eq1-2} and \eqref{matter-flow} take the same form, 
we need two additional equations to determine the whole evolution.
The necessity of an ansatz that offsets the surplus equations offers
physical insight into the dynamical role of the binormal vectors.

\vspace{.2cm}

Once we determine these two additional equations, it also fixes the binormal vectors $Q^\perp_a$ and $\tilde Q^\perp_a$.
The choice of any of these two auxiliary field fixes the whole equation of motions.
In retrospect, we can see that previous articles had their own choice for $Q_a^\perp$. 
In Ref.~\cite{Andersson2011}, Lopez-Monsalvo \& Andersson had chosen $Q_a^\perp =- \kappa \tilde{Q}_a^\perp$ and obtained an easy form of relativistic Cattaneo equation.
The Carter's choice~\cite{Carter89} for $Q_a^\perp$ corresponds that the binormal part of the force vanishes: $f^{0\perp}=0$ i.e., $Q^\perp_a=- \kappa (1-\gamma) \tilde{Q}^\perp_a$.
His choice had appeared to make the master function $\Lambda$ integrable.
However,  the choice makes the thermal system  unstable~\cite{Olson1990}.
Note that one cannot take both equalities because it over-determines the variables (10 equations for 8-variables) unless $q=0$. 
A lesson from these results is that the choice of the auxiliary fields is not a preference but a physically relevant one. 
Therefore, it is essential to have a good principle in the choice.

Noting Eq.~\eqref{d mu eq}, we have a better chance of choosing the auxiliary fields. 
The equation is quite interesting because it does not contain the heat flow $q^a$ directly and describes the behaviors of the thermodynamic quantities binormal to the heat and the matter flows. 
Since heat does not propagate along with those directions, we can say that two adjacent subsystems along these directions are in (thermal) equilibrium.
Let us search for ways to constrain the thermodynamic system to compensate for the two missing pieces of information. 
As we have described in the previous subsection, systems in thermal equilibrium satisfy the relations from Eq.~\eqref{dn} to Eq.~\eqref{Klein}.
Because the first two are relations for scalar quantities, we cannot use them to constrain the physical properties for the binormal directions. 
On the other hand, one may require one of the Tolman temperature gradient, and the Klein relation could hold between two subsystems but not both because we have only two remaining freedoms.  
Now we have three options:
\begin{enumerate}
\item {\it Tolman-like ansatz:} Choose to impose the Tolman temperature gradient to hold:
\be{Tolman local}
\perp_a^c \mathcal{T}_c =0.
\ee
This choice is natural because the formal definition of thermal equilibrium leads to  $\gamma_a^c\mathcal{T}_c=0$. In the absence of gravity, this choice implies that ``when no heat flows along a particular direction, two adjacent systems along that direction have the same temperature and vice versa.'' 
In a sense, this is a natural extension of the concept of thermal equilibrium in the presence of heat flow.

\item {\it Klein-like ansatz:} Choose to impose the Klein relation:
\be{Klein local}
\perp_a^c (\nabla_c \mu) =0.
\ee
This choice is one of the simplest. 
However,  the Klein relation must be an accidental relation for the single number flow system~\cite{Kim:2021kou}.

\item {\it Mixed ansatz:} Choose to impose a mixed constraint with the form: 
\be{mixed local}
\perp_a^c \left[ \phi\beta \Theta (\nabla_c \mu)- \nu \mathcal{T}_c\right] =0 ,
\ee
where $\phi$ denotes an arbitrary function of thermodynamic variables which must be dependent on a theory at hand. 
When $\phi =1$, Eq.~\eqref{mixed local} presents a simple relation between the binormal vectors.
\end{enumerate}

Of course, there could be many other choices.
Especially one may also use the form $\perp_a^c \mathcal{T}_c \propto \perp_a^b [\ve{q} \cdot \bm{d\Theta}]_b$.
It is because the term on the right-hand side vanishes in thermal equilibrium.
We do not pursue those possibilities here. 
Even though we proceed in a general setting, for the time being, we use the first choice based on the Tolman temperature gradient when we need to consider a specific solution, especially for the linear level analysis.
Let us write down the binormal vectors in this choice explicitly: 
\bea \label {Qs2}
\frac{Q^\perp_a}{\kappa} + \tilde{Q}^\perp_a 
&=& 
-\perp_a^b\left(
     \frac{\beta}{\Theta} [\ve{q}\cdot \bm{d\Theta}]_b 
 	+ n\beta \Theta \nabla_b \mu \right) , \nn \\
\frac{Q^\perp_a}{\kappa} + (1-\gamma) \tilde{Q}^\perp_a  
&=&  
 \perp_a^b\left( \frac{\alpha}{\bar\sigma \Theta}  
 	[\ve{q}\cdot \bm{d\Theta}]_b 
-\frac{\sigma n\beta}{\bar\sigma} \Theta \nabla_b \mu\right).
\eea
Now, the binormal vectors are fully expressed in terms of the binormal parts of $[\ve{q}\cdot \bm{d\Theta}]_a$ and $\nabla_a\mu$.
In this case, the relativistic analog of Cattaneo equation~\eqref{eq1-2} and the heat flow equation~\eqref{matter-flow} loses the $\perp_a^c\mathcal{T}_c$ terms on the right-hand side.

\section{Linear Perturbations around a thermal equilibrium} \label{sec:5}
In this section, we study fluid states close to thermal equilibrium. 
Our goal is to determine whether the equilibrium states are stable or unstable under linear perturbations. 
We follow the analysis of Olson and Hiscock~\cite{Olson1990} and Andersson et.al.~\cite{Cesar2011}.
We first display an equilibrium state and then obtain the equations governing linear perturbations around the state.
Next, we examine the properties of acceptable solutions to these equations for the case of a homogeneous background equilibrium state.
Since physically acceptable perturbations admit spatial Fourier transforms on at least one spacelike surface, we study the properties of the exponential plane-wave solutions.  

We denote the difference by $\delta F$ between the actual non-equilibrium value of a field $F$ at a given spacetime point and the value of $F$ in the background, a fiducial equilibrium state with $\vartheta=0$. 
We use the notation $F(n,s,\vartheta)$ and $F\equiv F(n,s,\vartheta=0)$ to denote the non-equilibrium value of the field and the equilibrium value, respectively, if necessary.
Fields that do not include the prefix $\delta$ (e.g., $n,\rho,u^a, \cdots$) will henceforth refer to the fiducial equilibrium state.

\subsection{Linear perturbations}

The physical parameters are functions of $n$, $s$, and $\vartheta$. 
Hereafter, we prefer to use $\sigma = s/n$ instead of $s$.  
Hence, the changes to near-equilibrium values will be 
$(n,\sigma, \vartheta=0)\to (n+\delta n, \sigma+\delta\sigma,\delta \vartheta=\vartheta)$ and $u^a \to u^a +\delta u^a$, respectively. 
The correction terms are functions of $n$, $\sigma$, and $\vartheta$: 
\be{series}
\delta u^a = \delta u^a(n,\sigma,\vartheta) , \qquad 
\delta n= \delta n(n,\sigma,\vartheta), \qquad 
\delta \sigma =  \delta \sigma(n,\sigma,\vartheta).
\ee
However, because $\vartheta_a = \beta(n,s,q) q_a$, we have $\delta \vartheta = \beta (n,s,q) \delta q$ at least in the linear order.
Therefore, we may use $q$ rather than $\vartheta$  in the linear order without loss of generality.

We denote the general value of the number and the entropy densities by $n(n,\sigma,q)$ and $s(n,\sigma,q)$, 
where $n$ and $\sigma$ inside the parenthesis denote their equilibrium values when $q=0$.
These notations raise no confusion in the calculation hereafter.
In general, $\delta n$ and $\delta \sigma$ are independent variations from $\delta q$.
When one calculates the first order variation, we use, for example, 
\bea \label{rho:q}
\delta \rho(n,\sigma,q) &\equiv& \rho(n+\delta n,\sigma+ \delta \sigma,q) - \rho(n,\sigma) ; \nn \\
 &=& \left(\frac{\partial \rho}{\partial n}\right)_{\sigma,q} \delta n
	+\left(\frac{\partial \rho}{\partial \sigma}\right)_{n,q} \delta \sigma
	+{\left(\frac{\partial \rho}{\partial q}\right)_{n,\sigma} \delta q}\nn\\
	&=&(\chi+\Theta \sigma)\delta n +n \Theta \delta \sigma,
\eea
where the equality in the third line 
uses the fact that $(\partial \rho/\partial q)$ is already a first order in $q$ as in Eq.~\eqref{conj}.
Therefore, $\delta \rho$ is effectively a function of $n$ and $\sigma$ to the first order.
Because of this result, $\delta\Theta$ and $\delta \chi$ are  also functions of $n$ and $\sigma$ only to the first order.

Now, we write down the differential equations of the linear fluctuations one by one.
 From the construction, $(u^a+ \delta u^a) (u_a+ \delta u_a)=-1$, $q_a \perp (u^a + \delta u^a)$, $u^a \dot u_a=0$, and $u^a q_a =0$ with $q^a = \delta q^a$, we have the following relations:
\be{1st}
u^a \delta u_a =0 = u^a \delta q_a, \qquad u^a \dot \delta u_a+ \delta u_a \dot u_a =0 .
\ee
This equation implies that $\delta u_a$ and $\delta q_a$ are normal to $u^a$ by nature.
Because of this constraint, there are 8-independent parameters, 
$\delta n$, $\delta\sigma$, $\delta u_a$, and $\delta q_a$, to be determined.

\begin{enumerate}
\item The number conservation equation $\nabla_a (n^a+\delta n^a) =0$ becomes
\be{delta n}
 (\nabla_a n) \delta u^a + n (\nabla_a \delta u^a ) + \frac{d}{d\tau}\delta n + \theta \delta n=0,
\ee
where we use  $\theta \equiv \nabla_a u^a = -\dot n/n $ in Eq.~\eqref{dn}.

\item Because the right-hand side of Eq.~\eqref{2nd law} is quadratic in $q$, the creation rate of the entropy vanishes, $\nabla_a(s^a +\delta s^a) =0$, to the linear order in $q$.
This equation becomes
\bea
\frac{d \delta (n\sigma)}{d\tau}+ \theta \delta (n\sigma)  
    +   \nabla_a( n\sigma) \delta u^a 
	+ n \sigma (\nabla_a \delta u^a) +\nabla_a \left(\frac{q^a}{\Theta}\right) =0.
\eea
Here, $\delta (n\sigma)$ means $(\delta n) \sigma + n (\delta \sigma)$.

\item
The relativistic analog of Cattaneo equation~\eqref{eq1-2} becomes, to the linear order in $q$, 
\bea \label{eq1-1st}
&&\frac{\delta q_a}{\tilde \kappa}
+\beta \left[ \dot q_a + (\nabla_a u_b) q^b 
		+2\perp_a^b  q^c \nabla_{[c} u_{b]} 
	\right]  \nn \\
&& \qquad	 =-n\beta[(\mu+\sigma) \dot \Theta + \Theta\dot \mu ]\delta u_a 
- \frac{q_a q^b}{q^2}\delta\mathcal{T}_b -\perp_a^c n\beta [(\mu+\sigma)\delta \mathcal{T}_b+ \Theta \nabla_b \delta \mu]   .
\eea
Here, $\tilde\kappa=\kappa$ for an equilibrium system satisfying $\gamma_n=0$ and $\dot \beta=0$. Here, we introduce an abbreviation
\be{d T}
\delta \mathcal{T}_b \equiv \nabla_b \delta \Theta + (\delta \Theta) \dot u_b+ 
	\Theta \delta \dot u_b .
\ee

\item The matter-flow equation~\eqref{matter-flow} becomes, to the linear order in $q$, 
\bea \label{heat flow2n}
&&
\frac{\sigma q_a}{\kappa'}
 -\alpha \left[ \dot q_a
 + \nabla_a u_b q^b 
+ 2\perp_a^b  q^c \nabla_{[c} u_{b]}
	\right] \nn \\
&& \qquad\qquad
=n\alpha[\dot \chi+\sigma \dot\Theta ] \delta u_a
+\frac{q_a q^b}{q^2} (\mu \delta\mathcal{T}_b+\Theta \nabla_b \delta \mu)  
+\perp_a^b n \alpha \left[ (\mu+\sigma) \delta\mathcal{T}_b  + \Theta  \nabla_b \delta\mu  \right] 
 ,	
\eea
where we use $\bar \sigma \approx \sigma$.
Here, $\kappa' =\kappa$ for the equilibrium system with $\dot \alpha =0$. 
The projection of this equation to the binormal direction is  the same as that of Eq.~\eqref{eq1-1st}.
\end{enumerate}

\subsection{Stability analysis for fluids in a flat background}
We now consider solutions of the perturbation equations for a general first-order theory subject to the following restrictions.
We assume that the background is a flat Minkowski spacetime, $g_{ab} = \beta_{ab}$.
Naturally, the background equilibrium state is homogeneous in spacetime, and all background field variables have vanishing gradients. 
In the equilibrium background state, the fluid is at rest so that 
\be{u t}
u^a \partial_a = \partial_t \quad \rightarrow \quad \theta =0, \quad \nabla_a u_b=0 .
\ee 
This background gives 
\be{background}
\dot \rho =0 = \dot \Psi = \dot \Theta = \dot \chi = \dot \mu , \qquad \dot u_a =0= D_a \Theta = D_a \mu = D_a \beta = D_a \rho = D_a \Psi \quad \rightarrow \quad D_a n=0= D_a \sigma,
\ee
where we use the fact that $\Theta$, $\chi$, and $\mu$ are functions of $n $ and $\sigma$.
$D_a$ denotes the derivative operator associated with $\gamma_a^b$. That is, $D_a\varphi=\gamma_a^b\nabla_b\varphi$ for a scalar field $\varphi$.

We look for exponential plane-wave solutions to the perturbation equations, 
\be{delta F}
\delta F = \delta F_0 \exp(i k x + \Gamma t),
\ee
where $\delta F_0$ is constant, and $t$ and $x$ are two of the orthonormal coordinates on Minkowski space. 
Let the perturbations vary along the $x$-direction.
Then, we have $8$-independent physical parameters:  $\delta n$, 
$\delta s$, $\delta u_a = (0, \delta u_i)$, $\delta q_a = (0, q_i)$ with $i=1,2,3$.

The differential equations for the linear fluctuations are:
\begin{enumerate}
\item The number and the entropy conservation equations become
\be{n:1st}
	\delta u^1=i\Omega \frac{\delta n}{n},
\ee
where $\Omega\equiv \Gamma/k$ and
\be{s:1st}
q_1=i\Omega \Theta (n\delta\sigma).
\ee

\item The linear perturbations of the relativistic analog of Cattaneo equation~\eqref{eq1-1st} present three independent equations. 
The $q^a$ direction equation, after multiplying Cattaneo equation by $q^a$ and using Eqs.~\eqref{n:1st} and \eqref{s:1st}, becomes
\be{cat q1}
q^i\delta u_i= n \delta \sigma \delta \Theta -\left( \beta +\frac1{\kappa \Gamma}\right) \frac{q^2}{\Theta}  .
\ee
The equation for the binormal directions, multiplying $\perp_a^b$, presents two independent equations, 
\be{q perp1}
\delta u_j 
= \left[  (q^i\delta u_i) 
	+ \frac{i  q_1}{\Omega} 
	\left(\frac{\delta \Theta}{\Theta} 
		+ \frac{ \delta \mu}{\mu+\sigma}	
	\right)
   \right]\frac{q_j}{q^2}, 
 \qquad j=2,3.
\ee
When $q_1=0$, the equation~\eqref{eq1-1st} presents an additional equation, 
\be{q=0 1}
\frac{\delta \Theta}{\Theta} +\frac{\delta \mu}{\mu+\sigma} -i\Omega \delta u_1 =0.
\ee

\item As mentioned in the previous section, the matter-flow equation~\eqref{heat flow2n} presents only one independent equation from the $q^a$ direction. 
Writing the result, we get 
\be{heat flow2-3n}
\left( \frac{\mu+\sigma}{\kappa}-\nu \Gamma\right)\left(\frac{q}{\Theta}\right)^2
=- \Gamma n \delta \sigma \delta \mu.
\ee

\end{enumerate}

\subsection{The stability condition for the equilibrium system and ``Klein modes"}
Equation~\eqref{heat flow2-3n} seems to dictate the solution space divided into three cases: $\delta \mu =0$,  $\delta \sigma =0$, and $\delta \mu, \delta \sigma \neq 0.$
During the calculation, however, we find that the case with $\delta \sigma =0$ is a part of the case with $\delta \mu =0$, when the local equilibrium condition~\eqref{Tolman local} holds.
We will show this later in Eq.~\eqref{conj1}. 
Noting these facts, we divide the solution space into two cases: (A) $\delta \mu =0$ and (B) $\delta \mu \neq 0$.

\begin{enumerate}
\item[(A)] ``Klein (decaying) modes" ($\delta \mu =0$): \\
We take the name Klein because the Klein relation, $\gamma^a_b\nabla_a \mu=0$, keeps valid around the equilibrium configuration with the fluctuations of this type only.
First of all, from equation~\eqref{heat flow2-3n}, the value of $\Gamma$ is determined to be a real-negative number,
\be{result10}
\Gamma = \frac{\mu+\sigma}{\kappa\nu}. 
\ee
Because $\mu+\sigma$ and $\kappa$ are positive definite for ordinary matters, this imposes the stability condition for the modes to decay with time as
\be{stability}
\nu < 0. 
\ee
To the first order in $q$, by using $\nu = \alpha - \mu \beta$, $\mu = (\rho+\Psi)/n\Theta$, and
$$
\left(\frac{\partial \rho}{\partial \vartheta}\right)_{n,\sigma} = \frac{q}{\Theta} \quad \rightarrow \quad \beta = \frac{\Theta}{q} \left(\frac{\partial \rho}{\partial q}\right)_{n,\sigma} ,
$$
the stability condition  constrains the $q$-dependence of the energy density:
\be{stable cond}
\left(\frac{\partial \rho}{\partial q}\right)_{n,\sigma}  > \frac{q}{\rho+\Psi} \quad \rightarrow \quad \beta > \frac{\Theta}{\rho+\Psi} = \frac1{n(\mu+\sigma)}.
\ee
Therefore, if one series-expand the density as a function of $q$, 
 the stability condition restricts the quadratic part of the heat to form $\xi(n,s) q^2/2(\rho+\Psi)$ with $\xi > 1$. 
This condition is the same as Eq.~(48) in Ref.~\cite{Cesar2011}.
We may also write the decay rate $\Gamma$ to the form:
$$
\Gamma = -\frac{q}{\kappa\Theta}
	\left[ \left(\frac{\partial \rho}{\partial q}\right)_{n,\sigma} 
	-\frac{q}{\rho+\Psi}\right]^{-1} .
$$
Note that the negative definiteness of $\nu$ is crucial in obtaining a stable system.
This fact enlightens  why the Landau-Lifshitz model that requires $\nu=0$ fails to be a successful theory~\cite{Hiscock}.

Now, let us solve the equations for this case explicitly. 
For $q_1 \neq 0$, combining Eq.~\eqref{q perp1} with Eqs.~\eqref{n:1st}, \eqref{s:1st}, and \eqref{cat q1}, we get
\be{d mu = 0}
    \Omega^2 \left( \beta +\frac1{\kappa \Gamma} \right)(n \delta \sigma)
   + \frac{\delta \Theta}{\Theta} +\Omega^2 \frac{\delta n}{n} =0,
\ee
where $\beta +1/{\kappa \Gamma}= [n(\mu+\sigma)]^{-1}$ from Eq.~\eqref{result10}.
We can use the equation $\delta \mu=0$ itself  to present a relation between $\delta n$ and $\delta \sigma$. 
Given $\mu$ and $\Theta$ as functions of $n$ and $\sigma$, this equation presents a variational relation between $q$, $\delta n$, and $\delta \sigma$ in terms of thermodynamic quantities. 
Explicitly, we use Eq.~\eqref{delta T app} and \eqref{delta mu app}.
Then, $\delta \mu=0$ presents a constraint from Eq.~\eqref{delta mu app}:
\be{d mu=0} 
\left(c_s^2-\alpha_s\right)\frac{\delta n}n 
+\left(\frac{\alpha_s}{\mu+\sigma} -\frac{1}{c_v}\right)  \delta \sigma =0  . 
\ee
Using Eq.~\eqref{delta T app}, Eq.~\eqref{d mu = 0} becomes
\be{A eq2}
 \left( \alpha_s + \Omega^2\right)\frac{\delta n}{n} 
 + \left(\frac1{c_v}
 + \frac{ \Omega^2} {\mu+\sigma}
   \right) \delta \sigma =0 .
\ee
Now, combining Eqs.~\eqref{d mu=0} and \eqref{A eq2}, the value of the mode $k$ is determined by the second-order function in $k$:
\be{Omega:mu=0}
\left(\frac{c_s^2 - 2\alpha_s}{\mu+\sigma} +\frac1{c_v} \right)  \Omega^2 = \frac{\alpha_s^2}{\mu+\sigma} - \frac{c_s^2}{c_v}   ,
\ee
where $c_s$ denotes the adiabatic speed of sound defined in Eq.~\eqref{cs}.
Using the relation between the heat capacities for fixed volume and fixed pressure in Eq.~\eqref{diff cs} we get the wavenumber,
\be{modes}
k^2 = - \frac{c_p}{c_s^2} \left[\frac{(c_s - \alpha_s/c_s)^2}{\mu+\sigma}+\frac1{c_p} \right]\Gamma^2,
\ee
where $c_p$ denotes the heat capacity for fixed pressure. 
The square bracket on the right-hand side is positive definite.
Therefore, the wavenumber $k$ must be a pure imaginary number because $\Gamma$ is a real number~\eqref{result10}. 
The $k$ mode will have a decreasing profile with $x$.
In the present case, however the direction of the vector $q^a$ is still not determined.

Now, we determine how the modes behave.
From Eq.~\eqref{q perp1}, we readily find that the heat-flow vector along the binormal directions satisfies
\be{transverse}
\delta u_j = c q_j , \qquad j= 2,3,
\ee
where 
\be{c}
c \equiv  \left[  (q^i\delta u_i) 
	+ \frac{i  q_1}{\Omega} 
	\frac{\delta \Theta}{\Theta} 
   \right]\frac{1}{q^2}.
\ee
Here, we name the ratio $c$ between the variation of the particle path and the heat. 

Putting Eq.~\eqref{transverse} into Eq.~\eqref{c} and using Eqs.~\eqref{n:1st} and \eqref{s:1st}, we  get 
\bea
c = \frac{\delta u_1}{q_1} + \frac{i}{\Omega q_1} \frac{\delta \Theta}{\Theta}
	&=&
- \frac{1}{n \Theta (\mu +\sigma)} = - \frac{1}{\rho+\Psi},
\eea
where we use Eq.~\eqref{modes} with $\Omega = \Gamma/k$.
When $q_1 =0$, Eq.~\eqref{c} fails to determine $c$. 
However, one may get the same result through Eq.~\eqref{cat q1}.
Because $\rho+\Psi$ is positive for ordinary matter satisfying energy conditions, we generally have that the variation $\delta u_j$ directs the opposite direction to the heat. 
Finally,  Eq.~\eqref{cat q1} presents an identity.

Let us examine a few specific cases. \\
i) The case with $q_1=0$, $q_2, q_3\neq 0$ describes transverse modes.
In general, this transverse modes do not belong to ``Klein modes". 
Therefore, there may exist transverse modes which satisfy $\delta \mu \neq 0$. 
However, in subsequent subsection~\ref{sec:VC}, we show that the transverse mode automatically satisfies $\delta \mu=0$ when the Tolman temperature gradient ansatz~\eqref{Tolman local} holds along the binormal directions.
Therefore, we do not need to analyze the transverse mode separately.
Because $\delta \sigma =0$ in this case, we get $c_s^2=\alpha_s$ from Eq.~\eqref{d mu=0}.
Then, we naturally have, from Eq.~\eqref{Omega:mu=0},
$$
\Omega=\pm i c_s.
$$
Here,  $c_s$ represents the adiabatic speed of sound.
Because $\Gamma$ is real-valued, the $k= \mp i \Gamma/c_s$ mode presents an exponential profile along $x$.

ii) When $q_2=0=q_3$, with $q_1\neq 0$, the set of equations \eqref{d mu = 0} exactly reproduces that of the equation for the longitudinal modes (See description around Eq.~\eqref{eq} in the subsequent subsection.) with $\delta \mu =0$ in Eq.~\eqref{q perp4n} below.
The only difference from the pure longitudinal modes is the value of $\Gamma$ is determined by Eq.~\eqref{result10} in the present case, implying a pure decaying mode. 

iii) General modes belonging to this class do not require any  $q_i$ components to vanish. 
In this sense,  ``Klein modes" represent  mixed ones between the longitudinal and the transverse modes.

\item[(B)] When $\delta \mu, \delta \sigma \neq 0$, we have a bit messy equation,
\be{q perp3-1}
   \frac{\delta \mu}{\mu+\sigma} \left[1+\Omega^2 \frac{\Theta^2}{q^2} (n\delta \sigma)^2\right]
   +  \frac{\delta \Theta}{\Theta} 
   + \Omega^2 \left[\frac{\delta n}{n} 
  +\left( \beta +\frac1{\kappa \Gamma} \right) (n \delta \sigma )\right]
=0.
\ee 
We get a relation between $\delta n$ and $\delta \sigma$ by putting Eq.~\eqref{heat flow2-3n} into Eq.~\eqref{q perp3-1}.
It is 
\be{q perp4n}
  \frac{\delta \Theta}{\Theta} 
   +\frac{\delta \mu}{\mu+\sigma} 
 + \Omega^2
      \left( \frac{\delta n}{n} +	
      	\frac{\delta \sigma}{\mu+\sigma}  
    \right)
=0 ,
\ee
where we use Eq.~\eqref{ABC}.
Using Eqs.~\eqref{delta T app} and \eqref{delta mu app}, we may get an equation that relates $\delta n$ with $\delta \sigma$.
To find the evolution of the thermodynamic system, we require one additional equation, which we  get in the subsequent subsection. 
\end{enumerate}

\subsection{The role of the Tolman-like ansatz and the longitudinal modes} \label{sec:VC}
The binormal projection of the matter-flow equation, as was mentioned in the previous section, fails to present a new relation but reproduces the same equation as Eq.~\eqref{q perp1}.
For the case with $\delta \mu, \delta \sigma \neq 0$, the equations of motion are incomplete to determine the whole evolution but require additional information. 

Let us demand Eq.~\eqref{Tolman local}  so that the Tolman temperature gradient is satisfied with the directions in which the heat does not flow. 
We show how the requirement makes the theory complete. 
The variation of the ansatz~\eqref{Tolman local} gives 
\be{local equil}
\delta[\perp_c^a \mathcal{T}_b] 
=(\delta u_a) \dot \Theta + \perp_a^b \delta\mathcal{T}_b =0;
\qquad \delta\mathcal{T}_b= \nabla_b \delta \Theta + (\delta \Theta) \dot u_b+ 
	\Theta \frac{d(\delta u_b)}{d\tau}  .
\ee. 
Then, the linear perturbation~\eqref{delta F} around the flat background~\eqref{background} is required to satisfy
\be{conj1}
\delta u_j 
= \left[  (q^a\delta u_a) 
	+ \frac{i q_1 }{\Omega} \frac{\delta \Theta}{\Theta}  
   \right]\frac{q_j}{q^2}, 
 \qquad j=2,3.
\ee
When $q_1=0$, it additionally requires 
\be{q1=0 2}
ik \frac{\delta \Theta}\Theta +\Gamma \delta u_1=0.
\ee
For $\delta \sigma \neq 0$, combining Eq.~\eqref{conj1} with Eqs.~\eqref{n:1st}, \eqref{s:1st}, and \eqref{cat q1}, we get
\be{equil 3}
    \Omega^2 \left( \beta +\frac1{\kappa \Gamma} \right)(n \delta \sigma)
   + \frac{\delta \Theta}{\Theta} +\Omega^2 \frac{\delta n}{n} =0.
\ee
Note that this equation takes  the same form as Eq.~\eqref{d mu = 0}.
However, one should be cautious of the fact that the value of $\Gamma$ is undetermined here. 
Now, we are ready to analyze the solution. 

\begin{enumerate}
\item[(A)] The ``Klein modes" are solved already in the previous subsection.
The new relation~\eqref{conj1}  constrains a mode with $q_i\neq 0,~i=1,2,3$ to the ``Klein modes" satisfying $\delta \mu=0$ by comparing it with Eq.~\eqref{q perp1}.
Considering the transverse modes, one may also get $\delta \mu = 0$ by comparing Eq.~\eqref{q=0 1} with Eq.~\eqref{q1=0 2}.
Therefore, the transverse modes are a part of the ``Klein modes" when the Tolman temperature gradient holds for the binormal directions.
In other words, the transverse modes do not modify the ratio $\mu$ between chemical potential and temperature. 
In addition to the transverse modes, other modes also belong to this class, which mixes the transverse modes and the longitudinal modes. 

\item[(B)] The longitudinal modes:\\
Comparing Eq.~\eqref{conj1} with Eq.~\eqref{q perp1}, we find that when $\delta \sigma \neq 0$ and $\delta \mu \neq 0$, nontrivial modes exist only when $q_2=0=q_3$ with $q=q_1 \neq 0$.
Because the perturbations are along the $x$-direction, they correspond to the longitudinal modes.
The equation describing their evolution are Eqs.~\eqref{q perp4n} and~\eqref{equil 3}.
Subtracting the two equations, we get
\be{eq}
\frac{\delta \mu}{\mu+\sigma} +\Omega^2\left[\frac{\nu}{\mu+\sigma} -\frac{1}{\kappa \Gamma} \right]n \delta \sigma =0.
\ee
Using Eqs.~\eqref{delta T app} and \eqref{delta mu app}, we can express the two equations with the following forms,
\bea \label{delta s2}
&&\left[\frac1{nc_v}+  \Omega^2 
	\left( \beta +\frac1{\kappa \Gamma} \right)
	\right](n \delta \sigma)
   +\left(\alpha_s+\Omega^2\right) \frac{\delta n}{n}
  =0 , \nn \\
&& \left[\frac1n\left(\frac{\alpha_s}{\mu+\sigma}-\frac{1}{c_v} \right) 
	+\Omega^2
		\left(\frac{\nu}{\mu+\sigma} 
		-\frac{1}{\kappa \Gamma} \right)
	\right]n \delta \sigma +(c_s^2-\alpha_s) \frac{\delta n}{n} 
=0.  
\eea

We can find a nontrivial solution  by solving the fourth order equation for $\Omega$,
\be{Gamma}
A\Omega^4 + B \Omega^3+ C \Omega^2 + D\Omega+ E=0, 
\ee
where the coefficients are
\bea
&&A = -\frac{\nu}{\mu+\sigma}, \quad
B =  \frac{1}{\kappa k}, \quad
C= \frac1{n(\mu+\sigma)}
\left[ \left(c_s -\frac{\alpha_s}{c_s}\right)^2
	+ \frac{\mu+\sigma}{c_p} - n\nu c_s^2  
	\right] , \nn \\
&&D = \frac{c_s^2}{ \kappa k }, \quad
E=- \frac1n \left[\frac{\alpha_s^2}{\mu+\sigma}- \frac{c_s^2}{c_v}\right] = \frac{c_s^2}{n c_p} >0.
\eea
Note that the coefficient $A$ and $C$ are positive when $\nu< 0$, which is nothing but  the stability condition of the ``Klein modes".
Evidently, the other coefficients $B,~D,$ and $E$ are also positive definite when the adiabatic speed of sound exists. 
Therefore, all the coefficients are positive definite without additional requirements.
The solutions $\Omega$ for the quartic equation~\eqref{Gamma} must be negative definite if it exists or at least a non-negative real solution does not exist.
This fact advertises the priority of the approach to those in literature, where additional constraints are necessary.
Since we know algebraic solutions of the quartic equations,
one can obtain the explicit forms of $\Omega$ in terms of these coefficients in principle.
Instead of presenting lengthy expressions, we restrict ourselves to checking the behavior of the solutions, especially for the dependency on $k$.

In the high $k$ limit, we may ignore $B$ and $D$. 
Note that $A X^2+C X +E=0$ has two negative real roots $X_\pm = (-C\pm \sqrt{C^2-4AE})/2A$ when $\nu < 0$  automatically because $A,~C,~E>0$ and
\bea
C^2- 4 AE 
&=&
\frac1{n^2(\mu+\sigma)^2}\left[\left(c_s -\frac{\alpha_s}{c_s}\right)^4+ 2\left(c_s -\frac{\alpha_s}{c_s}\right)^2\left( \frac{\mu+\sigma}{c_p} - n\nu c_s^2 \right)+\left( \frac{\mu+\sigma}{c_p} + n\nu c_s^2  	\right)^2 \right]\geq 0 .
\eea
Thus, $\Omega$ must have pure imaginary solutions $\pm i \sqrt{|X_\pm|} $ representing sound waves without additional requirements.
This fact also shows a difference from the results in the literature, in which  constraints for sound speed were usually added to make it nonnegative. 

To consider the finite $k$ effect, we set $\Omega =\pm i \sqrt{|X_\pm|} + w/k$ and find $w$ to get the $O(1/k)$ value in the short wavelength case. 
Then, we find
$$
w =  \frac1{\kappa}\frac{|X_\pm| - c_s^2}{2(2A|X_\pm| -C)}
	=  \frac1{\kappa}\frac{\mp(|X_\pm| - c_s^2)}{2 \sqrt{C^2-4AE}} . 
$$
For the system to be stable, the real part $w$ should be negative, which restricts the adiabatic sound velocity: 
\be{mode merger}
|X_-| \leq c_s^2 \leq |X_+| .
\ee
As noted in Ref.~\cite{Cesar2011}, this result is consistent with the notion that ``mode merger'' signals at the onset of instability~\cite{Samuelsson:2009up}.
Therefore, high $k$ modes should be stable with this condition satisfied.
Because there are two real roots, there exists the second sound. 
Note that the stability condition is far simply satisfied than those in the literature.
Especially no other conditions than $\nu <0$ and \eqref{mode merger} are required.

To ensure  causality, we need to require $|\Omega| \leq 1 \to -1\leq X < 0$.
This condition gives  $A- C +E > 0,$ and $2A> C$. 
Remember that $AX^2+CX+E=0$ always has two negative real roots.
The first condition presents
\be{cau1}
(1- c_s^2) \left(\frac{\mu+\sigma}{c_p} - n \nu \right) > \left(c_s-\frac{\alpha_s}{c_s}\right)^2 .
\ee
Because the right-hand side is non-negative, this inequality constrains the adiabatic velocity of sound to $c_s^2< 1$.
The term inside the parenthesis of the left-hand side is automatically positive definite because of Eq.~\eqref{stability}.
Once this is satisfied, it additionally constrains the value of $\nu$ to 
\be{cau1-1}
n \nu < \frac{\mu +\sigma}{c_p} - \frac{(c_s-\alpha_s/c_p)^2}{1-c_s^2}.
\ee
The second equation further constrains the value to 
\be{causality}
n\nu <  -\frac{\left(c_s - \frac{\alpha_s}{c_s}\right)^2
	+\frac{\mu+\sigma}{c_p} }{2-c_s^2} .
\ee
Interestingly, this inequality is identical to Eq.~(83) in Ref.~\cite{Cesar2011} even though the functions corresponding to $A$ and $C$ were different from ours.
Summarizing, the causality holds once the adiabatic speed of sound is slower than the light speed, and $\nu$ is small enough to satisfy the two inequalities~\eqref{cau1-1} and \eqref{causality}.
The combination of $2A> C$ with $C^2 > 4AE$ gives $A> E$ and $C> 2E$. 
These inequalities also present inequalities satisfied by $\nu$.
However, they must belong to part of Eq.~\eqref{causality} because $C^2> 4AE$ is an identity once $\nu < 0$.

For the case of a long wavelength, which presents the true-hydrodynamic limit, the solution takes the form, 
$$
\Omega \approx \pm i c_s 
	- \frac{1}{2n(\mu+\sigma)} \left(c_s - \frac{\alpha_s}{c_s}\right)^2\kappa k .
$$
This result is identical to Eq.~(66) in Ref~\cite{Cesar2011} or Eq.~(40) in Ref.~\cite{Hiscock1987}.
The analysis for the long-wavelength modes will proceed similarly, so we stop here.

\end{enumerate}

\section{Summary and Discussions} \label{sec:6}

We studied the problem of heat conduction in general relativity by using Carter's variational formulation. 
Especially, we have focused on the formal symmetry between the number and the caloric flows.
Even though we concentrate on a particular system consisting of only single fluid and a caloric flow, it bears many new features of a general multi-fluid system.
After scrutinizing the energy-momentum conservation law $\nabla_a T^{ab}=0$ in the directions of the caloric and the number flows, we wrote the creation rates of the entropy and the particle as combinations of the vorticities of temperature and chemical potential.
For this purpose, we formally hold the particle creation rate until we consider an equilibrium system. 

The rationale in this work is to follow the basic principle that heat produces entropy according to the second law~\eqref{2nd law}.
It leads us to obtain two {\it new} heat-flow equations, a relativistic analog of Cattaneo equation~\eqref{eq1} and a matter-flow equation~\eqref{eq2}.
The former is a generalization of the original Cattaneo equation, and the latter corresponds to the matter part of the force equations. 
We have incorporated two additional vectors, $Q^\perp_a$ or $\tilde{Q}^\perp_a$, into the formulation.
These vectors are binormal to the directions of the caloric flow and the number flow. 
This explicit representation makes us possible to discuss their dynamical roles,
although they are not associated with entropy production.

These binormal vectors provide a firm physical ground for making an ansatz in describing the whole evolution of the thermodynamic system.
Without them, making an ansatz is arbitrary or could lead to unphysical results.
Based on this consideration, we proposed a proper ansatz that reflects the core property of (local) thermal equilibrium. 
The ansatz states that the presence of heat does not generate the Tolman temperature gradient in the directions normal to the heat flow.
It is a natural choice because heat only affects physics in its flow direction.

As an application, we examined the linear stability of an equilibrium state in a flat background.
We discovered that there exist {\it new} ``Klein modes" which generalize the known transverse modes.
When $\delta \sigma$ vanishes, i.e., for fluctuations where the specific entropy does not change, the Klein modes become the same as the transverse mode.
However, the fluctuations with $\delta \sigma \neq 0$ mix the transverse modes with the longitudinal modes.
We explicitly obtain the condition for the stability and causality of the thermodynamic system. 
We find that the stability is crucially dependent on how the heat contributes to the energy density.
We show that the stability requirement is less stringent than that of the literature.

 In our analysis, we have studied the case when the covariant derivative of the tangent vector vanishes, $\nabla_a u_b=0$ in its equilibrium state.
However, in general,  $\nabla_au_b$ may not vanish for a curved  background spacetime, e.g.,  in an expanding universe.
Even when the background spacetime is flat,
it can be non-zero in a stationary rotating thermal equilibrium.
In these cases, we should consider the related coupling terms in the heat equation~\eqref{eq1-1st},
\bea
\beta \left[(\nabla_a u_b) q^b +2\perp_a^b  q^c \nabla_{[c}u_{b]} \right].
\eea
The first term has been shown in the literature,
whereas the second term is a new contribution from the binormal vectors $Q^\perp_a$ and $ \tilde{Q}^\perp_a$.
An example of a stationary rotating thermal equilibrium
can be a plasma state in nuclear fusion reactors. 
In the equilibrium, the expansion $\theta$ and 
the shear of the trajectory $\sigma_{ab}$ are nullified by electromagnetic fields,
whereas the twist $\omega_{ab}$ remains constant. 
In this case, the above coupling terms become
$$
\beta \left[\omega_{ab} q^b +2\perp_a^b  q^c \omega_{cb} \right]
$$ 
which shows the effects of the binormal vectors on the stability.

In this work, we have considered the ansatz that the thermodynamic system satisfies the Tolman relation locally in binormal directions. 
An alternative ansatz for the binormal vectors  that leads to another theory of heat conduction could be possible.
It is interesting to ask what happens in those cases.

The meaningful difference from previous works is conspicuous by the binormal vectors 
$Q_a^\perp$ and ${\tilde Q}_a^\perp$ 
since they directly affect the analysis by incorporating them into the heat-flow equations.
The creation rate of particles, $\Gamma_n$, may not vanish in general
in the presence of dissipation~\cite{Andersson:2013jga}.
Carter's formalism uses the vanishing property in deriving the Euler variations of the number flow. 
The dissipation may modify the formulation significantly, which will affect the heat-conduction equation too. 
It will be an interesting question for further study.

%
\section*{Acknowledgment}
This work was supported by the National Research Foundation of Korea grants funded by the Korea government NRF-2020R1A2C1009313.

\appendix

\section{Miscellaneous expressions}
For later convenience, we write the components of the force $\bm{f}^0$ in each direction.
By using the relativistic Cattaneo equation~\eqref{eq1}, we separate the force $\bm{f}^0$ into the $u_a$, $q_a$, and the binormal directions  to get
\be{f0 comp}
f_a^0 \equiv f^0_0 u_a +f^{0\shortparallel  } q_a + f^{0\perp}_a, 
\ee
where, 
\bea \label{f0 012}
f_0^0 &=& \Theta \Gamma_s-\left(\frac1{\kappa}+\frac{\Theta \chi \Gamma_n}{q^2} \right) (q^a s_a), \nn \\
f^{0 \shortparallel} &=&
\frac{q^a\Theta_a}{q^2}\Gamma_s-\left(\frac1{\kappa}+\frac{\Theta \chi \Gamma_n}{q^2} \right)s, \nn \\
f^{0\perp}
&=&  \frac{1}{ \Theta} \perp_a^c [\ve{q}\cdot (\bm{d \Theta})]_c   + \frac{s}{\kappa} (Q^\perp_a+\kappa \tilde{Q}_a^\perp)  .
\eea
Here, the orthogonality relations $ \ve{u}\cdot \bm{f}^{0\perp} =0 = \ve{q}\cdot \bm{f}^{0\perp} $ are apparent. 
We also separate the force $f^1_a$ for each direction by using Eq.~\eqref{eq2}:
\bea \label{f1 comp}
f^1_0 &=& \chi \Gamma_n , \nn \\
f^{1\shortparallel} &=&\frac{q^a\chi_a}{q^2}\Gamma_n
	+ n \bar{\sigma}
	\left(\frac1{\kappa} + (1-\gamma)\frac{ \Theta \chi \Gamma_n}{q^2} \right), \nn \\
f^{1\perp}_a &=&  -\frac{n\bar{\sigma}}{\kappa}  \big[ {Q}^\perp_a + \kappa(1-\gamma) \tilde{Q}_a^\perp\big]. 
\eea
Note that the orthogonal force $f_a^{1\perp}$ is thoroughly determined by the binormal terms ${Q}_a^\perp$ and ${\tilde Q}_a^\perp$. 
Explicitly, its contribution takes the same form as that in Eq.~\eqref{eq2}.
One can see that the temporal part of the conservation equation $\bm{f}^0+\bm{f}^1=0$ directly reproduces the second law~\eqref{2nd law}. 
Note also that the explicit formula, $\bm{f}^1 = f^1_0 \bm{u} + f^{1\shortparallel} \bm{q}  + \bm{f}^{1\perp}$, reproduces the heat flow equation~\eqref{eq2}.
Here, we use the second equation in Eq.~\eqref{forces}.

Note that the term $\left[ \ve{q} \cdot \bm{d\Theta} \right]_c$ on the right-hand side  
\be{q d Theta}
\perp_a^c\left[ \ve{q} \cdot \bm{d\Theta} \right]_c 
=\perp_a^c \left[ 2\Theta q^b \nabla_{[b} u_{c]} 
   + 2\beta q^b \nabla_{[b} q_{c]} - q^2 \nabla_c \beta\right] 
\ee
contains the vorticities $\bm{du}$ of the matter, $\bm{dq}$ of the heat, and the gradient of $\beta$. 
All the terms contain the heat $q_a$.
Only the first term is linear in $q$, but others are nonlinear.

The covariant derivative $\nabla_a u_b$ is decomposed as
\bea
\nabla_au_b=\frac{1}{d}\theta g_{ab}+\sigma_{ab}+\omega_{ab}, \qquad
\theta\equiv \nabla_cu^c,
\qquad 
\sigma_{ab}\equiv \nabla_{(a}u_{b)}-\frac{1}{d}\theta g_{ab},
\qquad 
\omega_{ab}\equiv \nabla_{[a}u_{b]} , 
\eea
where $\theta,\sigma_{ab}$ and $\omega_{ab}$ denote the expansion, shear and twist of the congruence of the trajectory and $d=2$ or 3 depending on matter, respectively.

\section{ Relation between the auxiliary fields and the sources of heat} \label{app:constrain}

To write the relativistic analog of Cattaneo equation, we first calculate explicitly,
\be{u d Phi}
[\ve{u}\cdot \bm{d\Theta}]_a =2 u^b \nabla_{[b} (\Theta u_{a]} + \beta q_{a]}) 
=\gamma_a^c \mathcal{T}_c + \dot\beta q_a + \beta[\dot q_a + (\nabla_au_b)q^b] .
\ee
Therefore, the relativistic Cattaneo equation~\eqref{eq1} becomes 
\be{eq1-1}
\frac{q_a}{\tilde\kappa} 
    + \beta \left[ \dot q_a + (\nabla_a u_b) q^b \right]
= -\gamma_a^c\mathcal{T}_c
	+\frac{Q^\perp_a}{\kappa} + \tilde{Q}^\perp_a ,
\ee
where $\tilde \kappa$ is given in Eq.~\eqref{tildes}.
Projecting to the directions normal to $q^a$ by multiplying $\perp_a^c$, we get
\be{proj1}
\beta \perp_a^c\left[ \dot q_c + (\nabla_c u_b) q^b \right]
= -\perp_a^c \mathcal{T}_c
	+\frac{Q^\perp_c}{\kappa} + \tilde{Q}^\perp_c .
\ee

By using the replacement $\Theta \to \mu \Theta$, $\beta \to \alpha$ in Eq.~\eqref{u d Phi}, we get
$
[\ve{u}\cdot \bm{d\chi}]_a 
= \gamma_a^c [\Theta (\nabla_c \mu ) +\mu \mathcal{T}_c] +\frac{d(\alpha q_a)}{d\tau} +  (\nabla_au_b)\alpha q^b .
$
Then, the heat flow equation~\eqref{eq2} becomes
\be{eq2-2}
\frac{q_a}{\kappa'} 
	- \frac{\alpha}{\bar\sigma} 
		\left[ \dot q_a
		+  (\nabla_au_b) q^b 
		\right] 
= \frac{1}{\bar \sigma} 
\gamma_a^c [\Theta (\nabla_c \mu ) +\mu \mathcal{T}_c] 
 + \frac{Q^\perp_a}{\kappa}+(1-\gamma) \tilde{Q}^\perp_a ,
\ee
where $\kappa'$ is in Eq.~\eqref{kappa'}.
Projecting to the binormal direction by multiplying $\perp_a^c$, we get
\be{proj2}
-\alpha\perp_a^c  \left[  \dot q_c
		+  (\nabla_cu_b) q^b 
		\right] 
= 
\perp_a^c [\Theta (\nabla_c \mu ) +\mu \mathcal{T}_c] 
 +\bar\sigma \left(\frac{Q^\perp_a}{\kappa}+(1-\gamma) \tilde{Q}^\perp_a\right) . 
\ee
Equating the two projections ~\eqref{proj1} and \eqref{proj2} we get
\be{cond}
 \alpha \left(\frac{Q^\perp_c}{\kappa} + \tilde{Q}^\perp_c \right)
+\beta \bar\sigma \left( \frac{Q^\perp_a}{\kappa}
			+(1-\gamma) \tilde{Q}^\perp_a\right)
 = 
\perp_a^c \left[-\beta \Theta \nabla_c \mu 
	+\nu \mathcal{T}_c \right] ,
\ee
where we have used Eq.~\eqref{bar chi}.

\section{The thermodynamic variables} \label{app:var}

During the calculations, we have defined various thermodynamic variables. 
The variables are assumed to be those in thermal equilibrium. 
Therefore, they are assumed to be functions of the number density $n$ and the specific entropy $\sigma$.
We use the following definitions, for heat capacity for fixed volume and $\alpha_s$:
\be{hc1}
c_v = \Theta \left(\frac{\partial \sigma}{\partial \Theta}\right)_n, \qquad
\alpha_s = \frac{n}{\Theta} \left(\frac{\partial \Theta}{\partial n}\right)_\sigma .
\ee
From this we can write the variation of the temperature,
\be{delta T app}
\frac{\delta \Theta}{\Theta} \equiv  \frac{1}{\Theta}\left[ \left( \frac{\partial \Theta}{\partial n}\right)_\sigma \delta n 
	+ \left( \frac{\partial \Theta}{\partial \sigma} \right)_n \delta \sigma \right] 
	= \alpha_s \frac{\delta n}{n} + \frac1{c_v} \delta \sigma .
\ee

To write down the variation of the chemical potential, we need to define the followings.
The adiabatic speed of sound is
\be{cs}
c_s^2 =  \left(\frac{\partial \Psi}{\partial \rho }\right)_\sigma = \frac{n}{\rho +\Psi} \left(\frac{\partial \Psi}{\partial n }\right)_\sigma 
= \frac{1}{\Theta(\mu+\sigma)}\left(\frac{\partial \Psi}{\partial n }\right)_\sigma.
\ee
The difference between the inverses of heat capacity for fixed volume and that for fixed pressure is
\be{diff cs}
\frac1{c_v} - \frac1{c_p} = \frac{n^3}{\Theta(\rho+\Psi) c_s^2} \left(\frac{\partial \Theta}{\partial n} \right)_\sigma^{~2} = \frac{1}{(\mu+\sigma) } \frac{\alpha_s^2}{c_s^2} .
\ee
Now, by using the formula, $\chi = \frac{\rho + \Psi}{n} - \Theta \sigma$, we get
$$
\left(\frac{\partial \chi}{\partial n}\right)_\sigma = 
	\frac1n \left(\frac{\partial \Psi}{\partial n}\right)_\sigma
	- \left(\frac{\partial \Theta}{\partial n}\right)_\sigma \sigma 
= \frac{\rho +\Psi}{n^2} c_s^2
	- \frac{\Theta \sigma}{n} \alpha_s 
= \frac{\Theta}{n} \left[(\mu+\sigma) c_s^2 - \sigma \alpha_s \right].
$$
Using the formula $(\partial/\partial n)_\sigma = (\partial/\partial n )_s + \sigma (\partial /\partial s)_n$ and 
$(\partial/\partial n)_s = (\partial/\partial n )_\sigma - \frac{\sigma}{n} (\partial /\partial \sigma)_n$ we additionally get
$$
\left(\frac{\partial \chi}{\partial \sigma}\right)_n 
=n \left(\frac{\partial \Theta}{\partial n }\right)_s 
= n \left(\frac{\partial \Theta}{\partial n }\right)_\sigma 
 -\sigma \left(\frac{\partial \Theta}{\partial \sigma }\right)_n
 = \Theta \left(\alpha_s - \frac{\sigma}{c_v} \right) .
$$
Using these results and $\mu = \chi/\Theta$, we get 
\bea
\left(\frac{\partial\mu}{\partial n}\right)_{\sigma} &=& \frac1{\Theta} \left[ \left(\frac{\partial \chi}{\partial n} \right)_\sigma 
	- \mu \left(\frac{\partial \Theta}{\partial n} \right)_\sigma 
	\right]
	=  \frac{\mu+\sigma}{n} (c_s^2- \alpha_s)  ,\nn \\
\left(\frac{\partial \mu}{\partial\sigma}\right)_{n} &=& \frac1{\Theta} \left[ \left(\frac{\partial \chi}{\partial \sigma} \right)_n 
	- \mu \left(\frac{\partial \Theta}{\partial \sigma} \right)_n 
	\right]
= \alpha_s - \frac{\mu+\sigma}{c_v} .
\eea
From this we get the variational relation of $\mu$ in terms of thermodynamic variables:
\bea \label{delta mu app}
\frac{\delta \mu}{\mu+\sigma}&\equiv & \frac{1}{\mu+\sigma}\left[ \left( \frac{\partial \mu}{\partial n}\right)_\sigma \delta n 
	+ \left( \frac{\partial \mu}{\partial \sigma} \right)_n \delta \sigma \right] 
	=  (c_s^2-\alpha_s) \frac{\delta n}{n} + \left(\frac{\alpha_s}{\mu+\sigma}-\frac{1}{c_v} \right) \delta \sigma .
\eea


\end{document}